# Three-Dimensional Virtual Histology in Unprocessed Resected Tissues with Photoacoustic Remote Sensing (PARS) Microscopy and Optical Coherence Tomography

Benjamin R. Ecclestone[1,2], Zohreh Hosseinaee[1], Nima Abbasi[1], Kevan Bell[1,2], Deepak Dinakaran[2,3], John R. Mackey[2,3], Parsin Haji Reza[1,*]

1. PhotoMedicine Labs, Department of Systems Design Engineering, University of Waterloo, 200 University Ave W, Waterloo, ON, N2L 3G1, Canada
2. illumiSonics Inc., 22 King Street South, Suite 300, Waterloo, Ontario N2J 1N8, Canada
3. Cross Cancer Institute, Department of Oncology, University of Alberta, 116 St & 85 Ave, Edmonton, Alberta, T6G 2V1, Canada

*phajireza@uwaterloo.ca

**Abstract:** Histological images are critical in the diagnosis and treatment of cancers. Unfortunately, the current method for capturing these microscopy images require resource intensive tissue preparation that delays diagnosis for many days to a few weeks. To streamline this process, clinicians are limited to assessing small macroscopically representative subsets of tissues. Here, we present a combined photoacoustic remote sensing (PARS) microscope and swept source optical coherence tomography (SS-OCT) system designed to circumvent these diagnostic limitations. The proposed multimodal microscope provides label-free three-dimensional depth resolved virtual histology visualizations, capturing nuclear and extranuclear tissue morphology directly on thick unprocessed specimens. The capabilities of the proposed method are demonstrated directly in unprocessed formalin fixed resected tissues. Here, we present the first images of nuclear contrast in resected human tissues, and the first 3-dimensional visualization of subsurface nuclear morphology in resected Rattus tissues, captured with a non-contact photoacoustic system. Moreover, we present the first co-registered OCT and PARS images enabling direct histological assessment of unprocessed tissues. This work represents a vital step towards the development of a real-time histological imaging modality to circumvent the limitations of current histopathology techniques.

## Introduction

Histopathological imaging of tissues is a principal tool in the treatment of cancers. These visualizations provide clinicians with the subcellular tissue characteristics to determine intrinsic disease biology, disease progression, tumor grade, cancer classification and prognosis. Current histopathological assessment with bright-field transmission microscopy requires tissue preparation with immunohistochemical staining [1,2]. During this workflow, tissues are resected, preserved in formalin, embedded into a paraffin substrate, sectioned with a microtome, and fixed to a microscope slide. Finally, the tissue slides are stained with exogenous dyes such as hematoxylin and eosin (H&E) to provide contrast [1,2]. Hematoxylin dye reveals cell nuclei, while eosin highlights extranuclear structures of the surrounding tissues [2]. In total, generating these bright-field microscopy compatible preparations takes several days [1,2]. Due to these practical limitations of preparation time and tissue volumes, only a small subset of resected tissue undergoes this histological processing. For breast cancer, less than ~2% of the excised tissues are assessed with conventional histology [3]. Clinicians must then make diagnostic decisions from the available preparations. Thus, current clinical practices rely heavily on the acuity of the histological workflow. However, each sample preparation step introduces a potential for variability in the resulting tissue preparations. For example, the specimens chosen to undergo histopathology are selected via macroscopic visual inspection and palpation [4]. As a result, the chosen samples may not be representative of the morphology of the bulk specimen and may exclude the most informative tissues. Even within the selected subset of tissues, only limited volumes of nuclei are visualized. Capturing different regions of nuclear and tissue morphology requires repeated thin sectioning, where each slice captures a ~5 µm section of tissue that must be independently fixed to a slide and stained prior to analysis. These compounding challenges have the potential to worsen patient outcomes.

Ideally, 3-dimensional nuclear and extranuclear-tissue morphology could be visualized label-free directly on unprocessed resected samples. This would avoid the delays of tissue preparation, enabling near real-time histological analysis. Moreover, direct imaging of resected tissues would facilitate more thorough analysis of excised specimens, rather than a restricted subset of tissues. This would permit the most highly informative tissues to be submitted, where required, for further immunohistochemical analysis. A technology capable of imaging in this fashion could revolutionize the workflow of anatomic pathology.

Although several novel histological imaging techniques have emerged, no single technology provides label-free 3-dimensional histological imaging of bulk unprocessed resected tissue samples. Most technologies in this histological imaging space require exogenous dyes or tissue clearing to image within specimens. These techniques include microscopy with ultraviolet surface excitation (MUSE) [5], fluorescence microscopy [6,7], non-linear microscopy [8,9], and light sheet microscopy [10,11]. Objectively, there are few techniques which have achieved measurable success providing label-free histology like contrast in resected tissues. The most notable are stimulated Raman scattering microscopy[12,13], photoacoustic (PA) microscopy[14–19] and optical coherence tomography (OCT) [20-25]. Regarding these methods, stimulated Raman scattering microscopy has recently provided histological visualizations in resected tissues [12,13]. However, this was applied in a transmission mode architecture only compatible with thin translucent tissue sections [12,13]. Therefore, PA microscopy and OCT are the most promising means to achieve label-free imaging of thick tissue samples.

OCT visualizes endogenous optical scattering contrast using an interferometric detection architecture. The interferometric mechanism enables OCT to capture columns of depth resolved scattering visualizations within tissues in single acquisitions. Combined with spatial scanning, this allows for recovery of 3-dimensional volumes of subsurface scattering structures. When applied to histological imaging, OCT has been shown to visualize subsurface features ~2 millimeters into tissues [20]. In research settings, OCT has been used to identify and delineate tissue regions, characterize subsurface lesions, and assess biopsy locations [20–24]. While the majority of works in this field focus on macroscopic tissue imaging, some recent publications have proposed ultra-high-resolution OCT systems for assessing submicron scale tissue morphology [25]. Though these subcellular visualizations have been achieved, the diagnostic utility of OCT is limited as the optical scattering contrast does not permit sufficient chromophore specificity [26]. As a result, OCT cannot provide direct nuclear contrast within tissues. Without these biomolecule-specific visualizations, OCT cannot match the diagnostic detail provided by the current pathology standard of H&E staining [26].

PA microscopy modalities offer distinct advantages as they capture optical absorption contrast. PA offers chromophore specific visualizations by leveraging the unique optical absorption spectra of biomolecules. This has been applied to selectively image DNA/RNA, hemoglobin, melanin, lipids, collagen, and more [14,17,27–31]. These unique chromophore specific visualizations position PA as a potentially powerful tool for label-free histological imaging [14,15]. The main challenge hampering the clinical adoption of this technique is that PA microscopy is a hybrid opto-acoustic modality. Conventional PA systems require physical contact with samples to perform imaging. However, a revolutionary new non-contact PA modality, Photoacoustic Remote Sensing (PARS) microscopy, has emerged as a frontrunner in label-free imaging [15,16,18,19,30,32]. Unlike traditional photoacoustic systems, PARS uses an all-optical pump-probe architecture. In PARS, the pump generates photoacoustic signals, which the probe captures as back-reflected intensity modulations [30]. Previously this technique was successfully applied to histology-like imaging in unprocessed resected tissue specimens, thin tissue sections, and paraffin embedded tissue blocks [15,16,18,19,33]. In these works, PARS provided histological visualizations analogous to hematoxylin staining of cell nuclei by leveraging the UV absorption contrast of DNA [15,16,18,19,33]. More recently, PARS hematoxylin-like imaging has been extended to full H&E emulation[15,19,34]. These methods leveraged endogenous absorption [15,19] and scattering contrast [34] to capture extra-nuclear morphology to accompany the PARS nuclear visualizations. However, these methods cannot match the single acquisition volumetric imaging of extra-nuclear tissue structures provided by OCT.

Here, we present a conjoined PARS microscope and swept source OCT (SS-OCT) system for rapid 3-dimensional virtual histology in bulk unprocessed resected tissue specimens. We use the UV excitation PARS microscope to provide chromophore specific recovery of nuclear morphology. Concurrently, we use the SS-OCT to capture 3-dimensional volumetric images of tissue morphology. We then merge the co-registered PARS and OCT data to provide three-dimensional, histological, and structural tissue visualizations. While a combined PARS and OCT system has previously been proposed for ocular and vascular imaging[35,36], this is the first such system for histopathology. Applying the SS-OCT in resected Rattus tissues we recover volumetric images containing adipocytes, ducts, fascia layers and other tissue features. Applying the PARS system in unprocessed resected human and Rattus tissues for the first time we present high spatial resolution imaging of predominately nuclear structures. Within the PARS visualizations we can identify tissue boundaries, assess nuclear atypia, arrangement, organization, and density. Moreover, we leverage the optical sectioning capabilities of the PARS microscope to capture the first 3-dimensional volumetric images of subsurface nuclear morphology in unprocessed tissues. Together this system enables label-free non-contact assessment of nuclear structures directly on 3-dimensional visualizations of resected tissue specimens. These visualizations provide qualitatively similar diagnostic features to conventional H&E preparations. If employed in a clinical setting, the proposed technique could allow direct histology-like imaging of unprocessed resected tissue samples. Thus, the PARS-OCT system holds the potential to dramatically reduce current limitations in the histopathological processing of tissues. This could provide near-real time histological analysis, circumvent current diagnostic limitations, and greatly improve patient outcomes.

## Methods

*Imaging System Architecture*

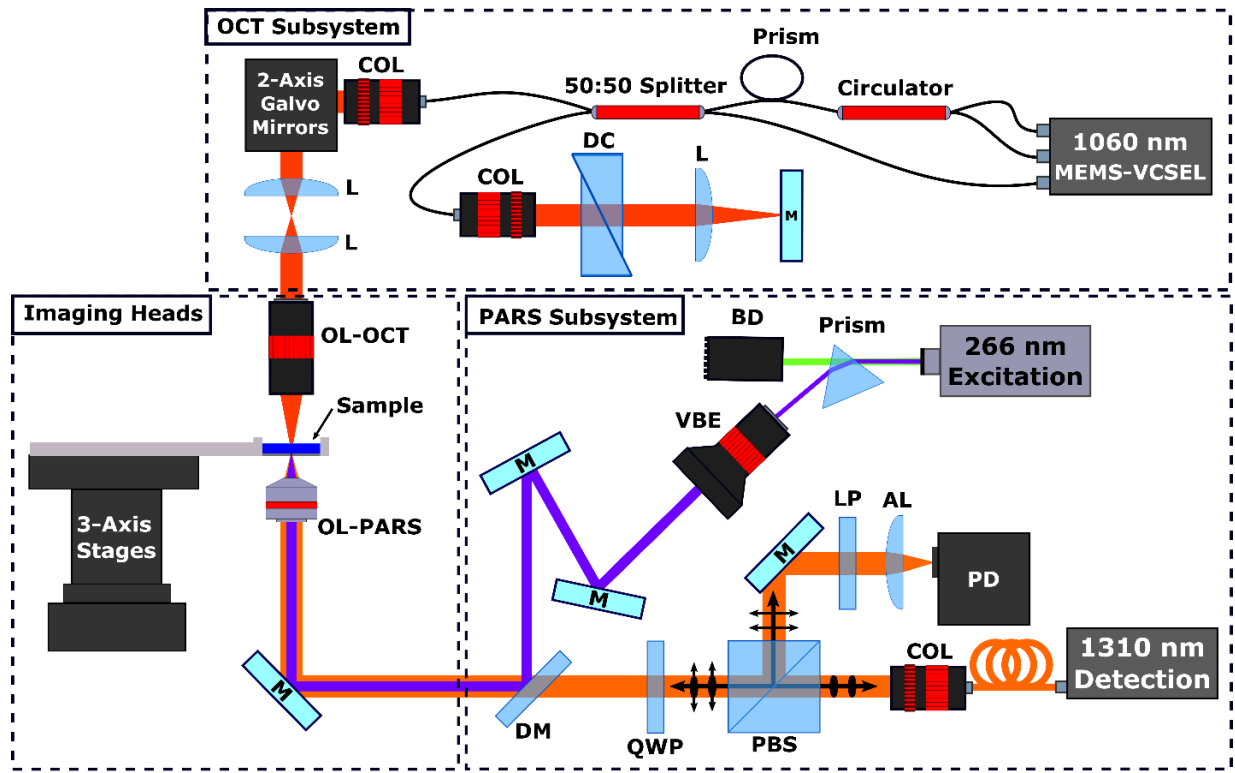

**Figure 1:** Simplified Schematic of the combined OCT and PARS system. Component labels are defined as follows: OCT objective lens (OL-OCT), PARS objective lens (OL-PARS), collimator (COL), dispersion compensation (DC), lens (L), beam dump (BD), variable beam expander (VBE), polarizing beam splitter (PBS), quarter wave plate (QWP), dichroic mirror (DM), long pass filter (LP), aspheric focal lens (AL), photodiode (PD), mirrors (M).

*OCT System*

The OCT system uses a MEMS-VSCEL light source (Thorlabs, Inc.) centered at ~1060 nm with ~100 nm full width sweep bandwidth and 60 kHz sweep rate. To synchronize data collection, A-line trigger, also known as wavelength sweep trigger, is supplied by the laser source. Concurrently, K-linear clock signals are provided by the Mach-Zehnder interferometer-based clock module integrated within the laser source. The swept source laser output is connected to a custom fiber optic interferometer. The interferometer consists of a circulator and a 50:50 fiber splitter to direct light into the reference and sample arms. The reference arm consists of, a pair of BK7 dispersion compensating prisms, and a translating mirror to set the zero delay. In the sample arm, collimated light is directed through a pair of two-axis galvo-scanner mirrors, a 1:1 telecentric pair, then focused on the sample using an infinity corrected microscope objective. Light returning from the sample is coupled back into the 50:50 splitter where it is combined with the reference beam. The combined beams are detected by the swept sources built-in dual balanced photodetector. The OCT signals from the photodetector are then recorded with a high-speed 16-bit digitizer card (ATS9351, Alazar Technologies Inc., Pointe-Claire, QC, Canada). Considering system performance, using a mirror, the maximum SNR of 105 dB was measured at ~ 100 µm away from the zero-delay line with incident power of ~ 1.5 mW. The SNR roll-off in free space was measured to be ~ 1 dB over a scanning range of 1.3 mm. SNR was calculated as a ratio of the maximum signal versus the standard deviation of the background.

*PARS System:*

For the PARS system, UV excitation is provided by a 266 nm 400 ps 50 kHz pulsed laser (WEDGE XF, Bright Solutions). The 266 nm excitation is separated from residual 532 nm output using a CAF2 prism. Once isolated, the UV excitation beam is expanded and combined with the probe beam. The 1310 nm probe is supplied by a continuous wave super-luminescent diode laser from Thorlabs (S5FC1018P, Thorlabs). The horizontally polarized probe beam is passed through a polarizing beam splitter and quarter wave plate into the imaging system. The combined excitation and probe beam are then focused onto the sample using a 0.5 NA reflective objective (LMM-15X-UVV, Thorlabs). The back-reflected probe beam containing the PARS signals return to the quarter wave plate and polarizing beam splitter by the same path as forward propagation. Upon passing through the quarter wave plate a second time, the returning probe beam becomes horizontally polarized. The horizontally polarized probe beam is redirected down the detection pathway by the polarizing beam splitter. Here, the probe beam is isolated through spectral filtering, then focused onto a photodiode (PDB425C-AC, Thorlabs).

*Imaging Reconstruction and Processing*

*OCT Image Reconstruction and Processing*

Raw OCT frames are captured by optically scanning the beam across the sample using the 2-axis galvo-mirrors. At each wavelength sweep trigger, the interferogram and corresponding location are recorded. To capture the spectral interferogram 2448 sampling points are acquired with the 16-bit digitizer, providing a depth range of ~12 mm. Directly following acquisition, the OCT reference spectrum is subtracted from the interferogram to remove DC bias. The data is then Fourier transformed, to extract the depth-resolved OCT signal. The top half of the Fourier transformed data is considered as valid for further processing. To generate volumetric data sets, series of A-lines (usually 500 to 800) were acquired to form a cross sectional B-scan. Series of adjacent B-scans (usually 500 to 800) are collected to form a volumetric data set. Once the entire volume is captured, the raw OCT data is numerically dispersion compensated up to the 5th order with a custom MATLAB algorithm. No additional post-processing was applied to the OCT results presented in this paper. Volumetric and enface images were then generated from the 3D data sets with ImageJ [37].

*PARS Image Reconstruction and Processing*

To form PARS images the co-focused excitation and probe beams are mechanically scanned across the sample. During scanning the excitation laser is pulsed continuously, interrogating regularly spaced points of absorption contrast. The lateral spacing of the interrogation points is modified by adjusting the stages mechanical scanning pattern and speed. At each interrogation event, the PARS signal, and interrogation location are recorded. These signals are captured using a high bandwidth 14-bit digitizer (RZE-004-300, Gage Applied). Location signals are recorded directly from the mechanical stages, while PARS signals are recorded by capturing a ~500 ns segment of photodiode signal.

The digitized signals are then streamed directly to the computer memory. Processing is applied to compress the time domain data, to a single characteristic PARS amplitude and corresponding location signal. Once scanning is complete, the scattered data is reconstructed into an image. Here, the location signals are used to impose the PARS signals onto a cartesian grid, forming a raw PARS image. Raw data is gaussian filtered to reduce scanning noise, then normalized and log scaled to convert the PARS signals to a decibel scale. To generate the final image, the logarithmic PARS image is rescaled based on histogram distribution to reduce background noise.

*PARS and OCT Image Co-registration*

Combined PARS and OCT visualizations were generated using the final PARS images, and processed OCT volumetric data. Prior to processing, the PARS images were overlayed over the max amplitude projection of the OCT volume to manually tune the co-registration, as there were some minor motion artifacts. Since the PARS images were substantially higher resolution than the OCT frames, the datasets had to be rescaled prior to combination. For the large area scans, the PARS image was compressed to match the OCT volume. Conversely, for the high-resolution frames, the OCT volume was interpolated to match the PARS image dimensions. Merging the PARS data into the OCT volume was performed one B-scan at a time. In each B-scan, the tissue surface was extracted based on the regional maximum of the OCT signal. The corresponding stripe of PARS contrast was then merged into the OCT B-scan along the extracted surface. The entire volumetric dataset was processed in this fashion. Volumetric and enface images were generated from the resulting 3D data sets with ImageJ[37].

*Sample Preparation*

Both human and Rattus mammary tissues were explored during this study. Human tissue samples were collected from a post-menopausal breast during a tumor excision surgery. These tissue samples were collected under protocols approved by the Research Ethics Board of Alberta (Protocol ID: HREBA.CC-18-0277) and the University of Waterloo Health Research Ethics Committee (Humans: #40275). Rattus tissue samples were collected from rats with naturally occurring mammary tumors. Rats found to have tumors were sacrificed according to the endpoint protocols laid out by the University of Waterloo Research Ethics committee (Animals: 41543). Following sacrifice, the mammary tumors were extracted.

In both cases, mammary tissues were resected and placed directly into a fixative solution of 10% formalin within 20 minutes of resection. Fixed tissues were transported directly to the PhotoMedicine labs at the University of Waterloo. Prior to imaging, the bulk tumors were sectioned to expose regions of dense tumor tissue. Then, the unstained bulk tissues were placed directly onto the PARS and OCT system, and imaged at room temperature. Following imaging, the samples were returned to their formalin storage containers.

**Results and Discussion**

To capture nuclear contrast, the PARS system targets the optical absorption peak of DNA using a 266 nm UV excitation. Applying PARS to unprocessed resected tissues we capture high-fidelity visualizations of nuclear structures, analogous to hematoxylin staining of cell nuclei. Such visualizations are shown in Figure 2. Here, bulk sections of formalin fixed tissues are placed directly onto the PARS imaging stage, then scanned with the PARS system. These images reveal nuclear structures directly within unprocessed formalin fixed Rattus tissues, and for the first time, unprocessed human breast tissues. Within the presented images, we can discern and differentiate clinically relevant diagnostic features of the nuclear morphology. For example, in Figure 2(a-i), we can differentiate between highly cellular regions and the surrounding nuclear sparse adipose and connective tissues. Observing a subsection of this large field image, Figure 2(a-ii) we can discern subtler diagnostic details. Within the higher magnification region, we can identify cellular structures organizing into a circular morphology indicating mammary glands and ducts. Looking at a smaller subsection of the large field, Figure 2(a-iii), highlights some of the gland and duct architecture and the boundary between the nuclear dense tissues (bottom right) and the sparse surrounding regions (top left). This small frame exemplifies the ~300 nm lateral resolution (measured on 200 nm gold nanoparticles) provided by the PARS system[18]. Within the small frame we are able to visualize individual cells, nuclei and how they are arranged in the tissue architecture. Applying this technique to human tissues for the first time, we achieve similar diagnostic quality (Figure 2(b) to (d)). Within the human tissues, Figure 2(b) we can discern multiple regions of dense nuclear structures, and the surrounding connective tissues. Looking closer at the hypercellular regions (Figure 2(c) and (d)), individual cells and nuclei become appreciable as well as their distribution within the tissue sample.

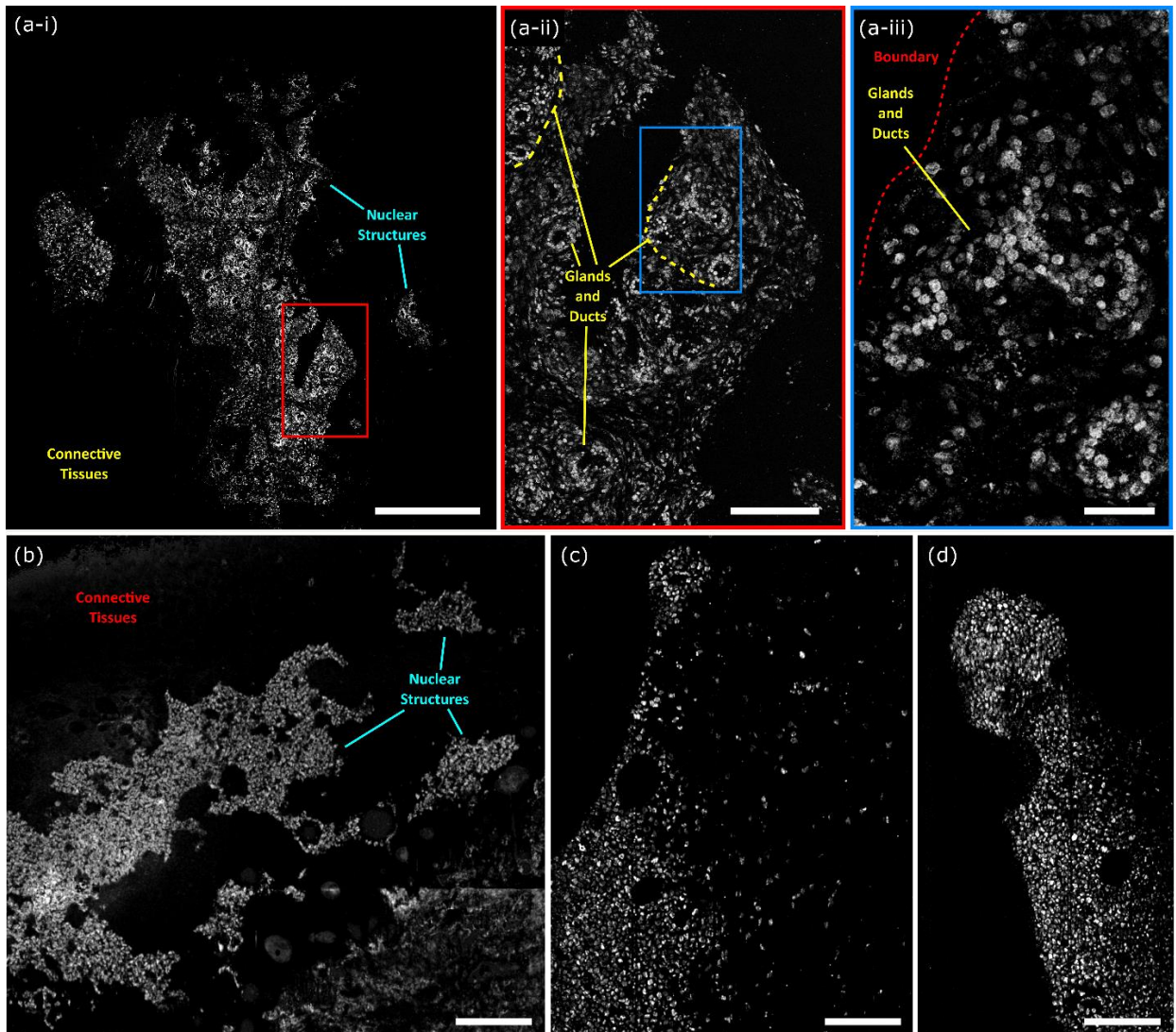

**Figure 2:** PARS histological imaging of nuclear morphology in unprocessed excised tumor tissues. (a) (i) Large field of view PARS image of resected Rattus mammary tumor. Scale Bar: 1 mm. (ii, iii) Subsections of the large field of view image in (i) highlighting regions of interest within the large field images. (ii) Scale Bar: 500 µm. (iii) Scale Bar: 25 µm. (b-d) PARS histological imaging of nuclear morphology in unprocessed formalin fixed excised post-menopausal human breast tissues. (b) Scale Bar: 275 µm (c) Scale Bar: 150 µm (d) Scale Bar: 150 µm.

In addition to high lateral resolution, the proposed PARS microscope provides high axial resolution. In this architecture, the PARS system provides a ~1.4 µm focal plane. Leveraging this tight axial focus, the PARS system may perform optical sectioning to capture isolated planes of depth resolved nuclear contrast within excised specimens (Figure 3 (a)). Here, we capture a series of nuclear visualizations from the sample surface through to ~30 µm in depth. This image series is then reconstructed to form a 3-dimensional representation of the subsurface nuclear morphology (Figure 3(b) through (d)). In this volumetric reconstruction, we resolve multiple overlapping layers of cellular morphology. Moreover, successively deeper layers in the 3-dimensional reconstruction shows subsurface cellular morphology of glandular breast tissue. These results represent the first report of a non-contact photoacoustic system capturing 3-dimensional volumetric images of subsurface nuclear morphology in resected tissues. Compared to the 5 µm thickness of current paraffin tissue preparations, PA based histopathology systems have been shown to image ~100 µm into resected tissue specimens [14,17]. As the UV excitation is highly absorbed and scattered within tissues, PA systems begin to rapidly lose contrast and resolution beyond this depth. However, even with this limitation the PARS virtual sectioning captures nuclear morphology equivalent to the depth provided by ~6 thin tissue preparations. Applied in a clinical setting, the PARS virtual sectioning capabilities would allow pathologists to rapidly assess subsurface nuclear structures without extensive sample processing.

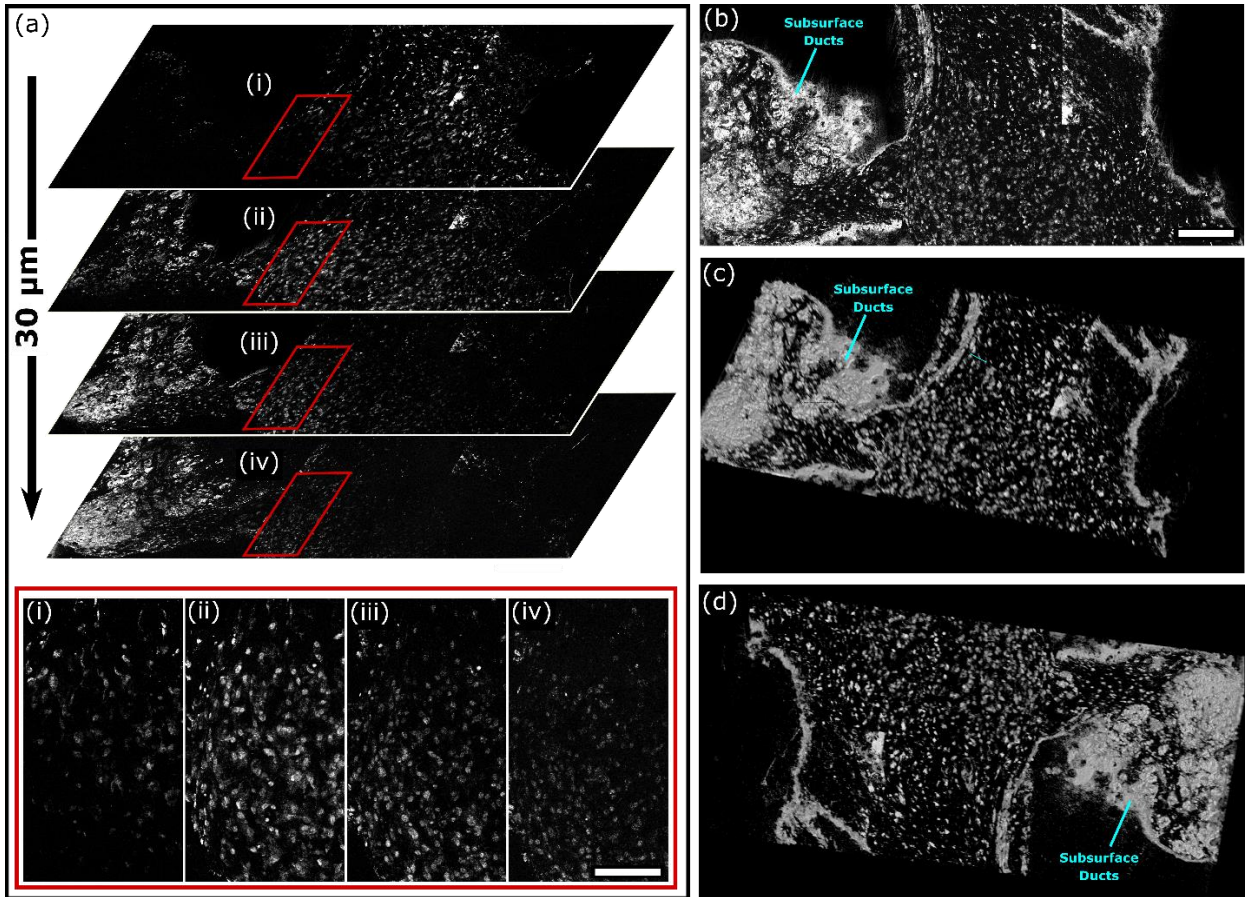

**Figure 3:** PARS 3-dimensional volumetric imaging of subcellular nuclear morphology in unprocessed excised Rattus mammary tumor tissue. (a) A subset of the optically sectioned volumetric image stack showing subsurface nuclear morphology from the surface to 30 µm into the tissue. Scale Bar: 125 µm. (b-d) Different angular views of the 3-dimensional reconstruction of the nuclei within the excised tissue sample. Scale Bar: 200 µm.

To compliment the nuclear morphology captured by the UV excitation PARS system, we apply OCT to image the bulk tissue structures, and subsurface specimen morphology. The proposed OCT subsystem provides 10.1 µm axial resolution in free space, corresponding to 7.3 µm in tissues, assuming an average refractive index of n = 1.38. Depending on the application, the system used either a 0.14 NA or 0.4 NA objective corresponding to ~15.6 µm and ~3.9 µm lateral resolution respectively. The lower 0.14 NA lens provides wide area grossing scans capturing nearly 1 cm$^2$, while the higher 0.4 NA lens provides high resolution imaging of ~0.5 mm$^2$ fields. In resected tissue specimens, the OCT system provides visualizations as presented in Figure 4. Large field scans capture entire tumor specimen, providing a 3-dimensional representation of the bulk tissue structure (Figure 4 (a)). Leveraging the unique depth resolved scattering contrast provided by OCT, we also assess the bulk specimens subsurface tissue morphology, Figure 4 (b-i) and (b-ii). Observing the subsurface structures, regions of breast connective and adipose tissue, fibrous structural tissue, and gland/duct tissue become evident. Changing over to the 0.4 NA lens, we capture smaller high-resolution visualizations of tissue structures. Several max projection images of small fields are shown in Figure 4 (c) through (e). Within these images, we can discern subtler features such as connective tissue bands segmenting structures within the tissue (4c), with potential spaces of ducts and glands interspersed. In other regions, we identify circular voids within tissues adjacent to bands of connective tissues (4d). Moreover, less feature rich regions are captured likely representing adipose tissue (4e).

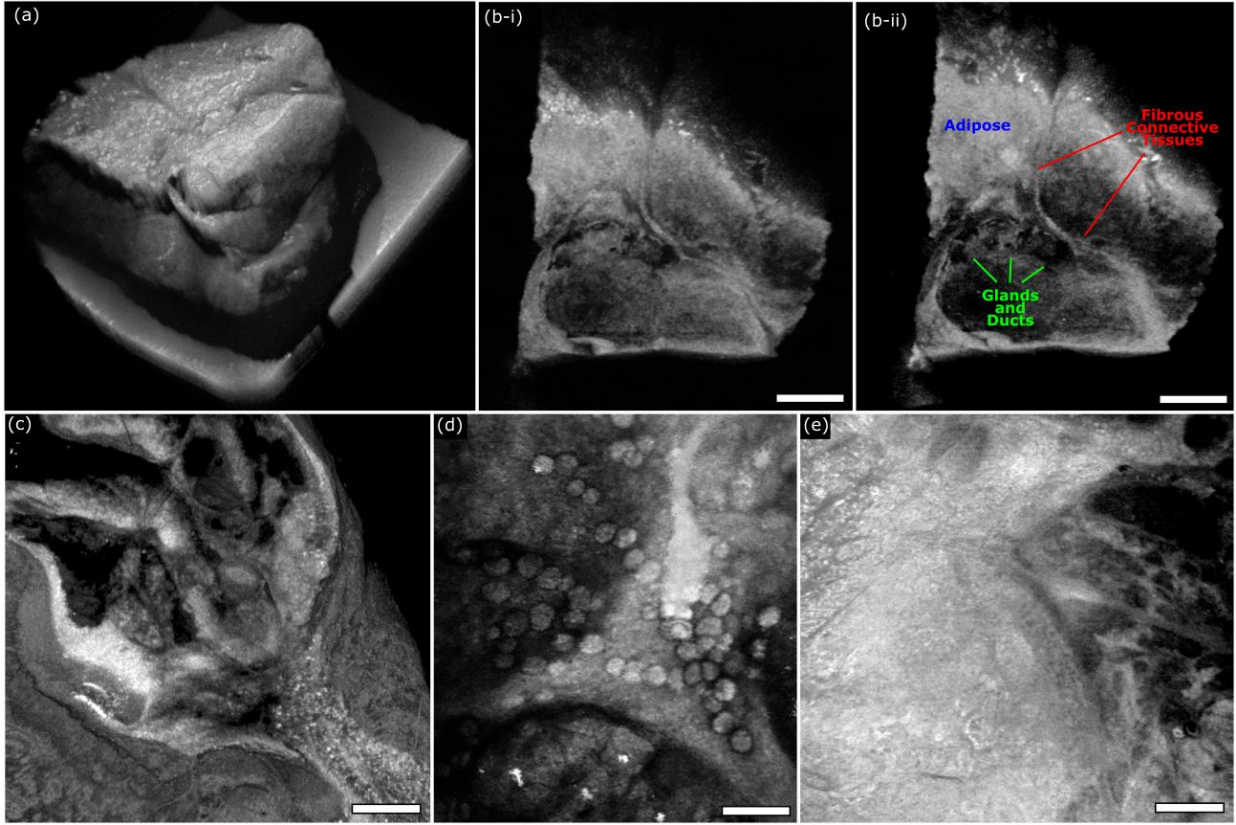

**Figure 4:** OCT imaging of unprocessed excised Rattus mammary tumor tissue. (a) Shows a large field OCT image capturing the surface morphology of an entire excised tumor specimen (b) Vertical sections showing subsurface tissue morphology of the tissue captured in (a). Scale Bar: 1.25 mm. (c-d) Small field maximum projection images of different regions of excised Rattus mammary tissues captured using the OCT system. (c) Scale Bar: 100 µm. (d) Scale Bar: 100 µm. (e) Scale Bar: 100 µm.

In isolation, PARS and OCT each present distinct advantages and weaknesses. The OCT provides unique depth resolved tissue morphology imaging; however, it cannot accurately highlight nuclear structures. Conversely, PARS provides chromophore specific visualizations readily capturing nuclear contrast, however, it cannot provide depth resolved tissue morphology in a single acquisition. Though, Bell *et al.* have proposed a coherence-gated PARS microscope which could provide single acquisition depth resolved chromophore specific tissue imaging in the future [38]. In the current implementation, we combine the PARS and OCT to leverage the advantages of each system. This dual modality PARS and OCT system provides a unique suite of clinically relevant imaging capabilities. To capture combined PARS and OCT images, tissues are first scanned with the PARS microscope Figure 5(a-i). The wide field PARS image captured here covers ~50% of the resected tissue specimen and highlights the nuclear structures of the tissue. Once PARS imaging is complete, tissues are imaged with the OCT system equipped with the 0.14 NA objective. The corresponding 3-dimensional OCT representation of the bulk tissue surface morphology is highlighted in Figure 5 (a-ii). We merge the co-registered datasets to provide PARS images of nuclear morphology directly on the 3-dimensional OCT tissue visualizations, Figure 5 (a-iii). These combined visualizations reveal the correlation between the PARS nuclear features, and the corresponding OCT based tissue morphology.

In practice, the PARS nuclear contrast is merged directly into the OCT volumetric data. Therefore, the combined PARS and OCT volumetric scans may be treated analogous to pure OCT images. This means 3-dimensional representations of tissues exhibiting nuclear morphology can be assessed from any angle and orientation, Figure 5 (b) and Figure 5 (c). Moreover, since the PARS data is added to the OCT volume, the subsurface tissue morphology may still be assessed as with traditional OCT volumes, Figure 5 (d). Limited rendering resolution is the main challenge associated with these 3-dimensional PARS and OCT visualizations. During data collection, the OCT scans are restricted to a resolution of ~800 by 800 pixels, due to the large volume of data. Therefore, in order to merge the PARS and OCT visualizations, the PARS data is down sampled to match the spatial resolution of the OCT. As a result, only the general cellular organization based on the nuclear contrast data can be resolved within the wide field representations. In order to view the nuclear morphology with higher resolution, subsections of interest from the large frame are re-imaged with the higher resolution 0.4 NA OCT. This allows us to produce high resolution visualizations of nuclear morphology directly on the OCT visualization Figure 5 (e). Within these subframes, (Figure 5 (e) to Figure 5 (h)), we can see areas of subcutaneous tissue demarked by more homogenous and smaller nuclei transitioning into more characteristic breast tissue architecture with duct and gland formation consisting of larger cells and nuclei (5e). Intervening adipose and connective tissue space is seen in Figure 5(f-g) between glandular breast tissue regions, with medium and small-sized blood vessels becoming evident (Figure 5(g)). Sparsely distributed nuclei within the connective and adipose tissue where individual nuclear atypia can be assessed within the PARS image (Figure 5(h)). This type of diagnostic interpretation can be ascertained more reliably with the context of both the high-resolution combined PARS and OCT renderings providing information (Figure 5 (e) to (h)). Applied in a clinical setting this would provide clinicians with visualizations of useful histopathology directly on excised tissues. This should expedite histopathological assessment by circumventing current sectioning and staining requirements.

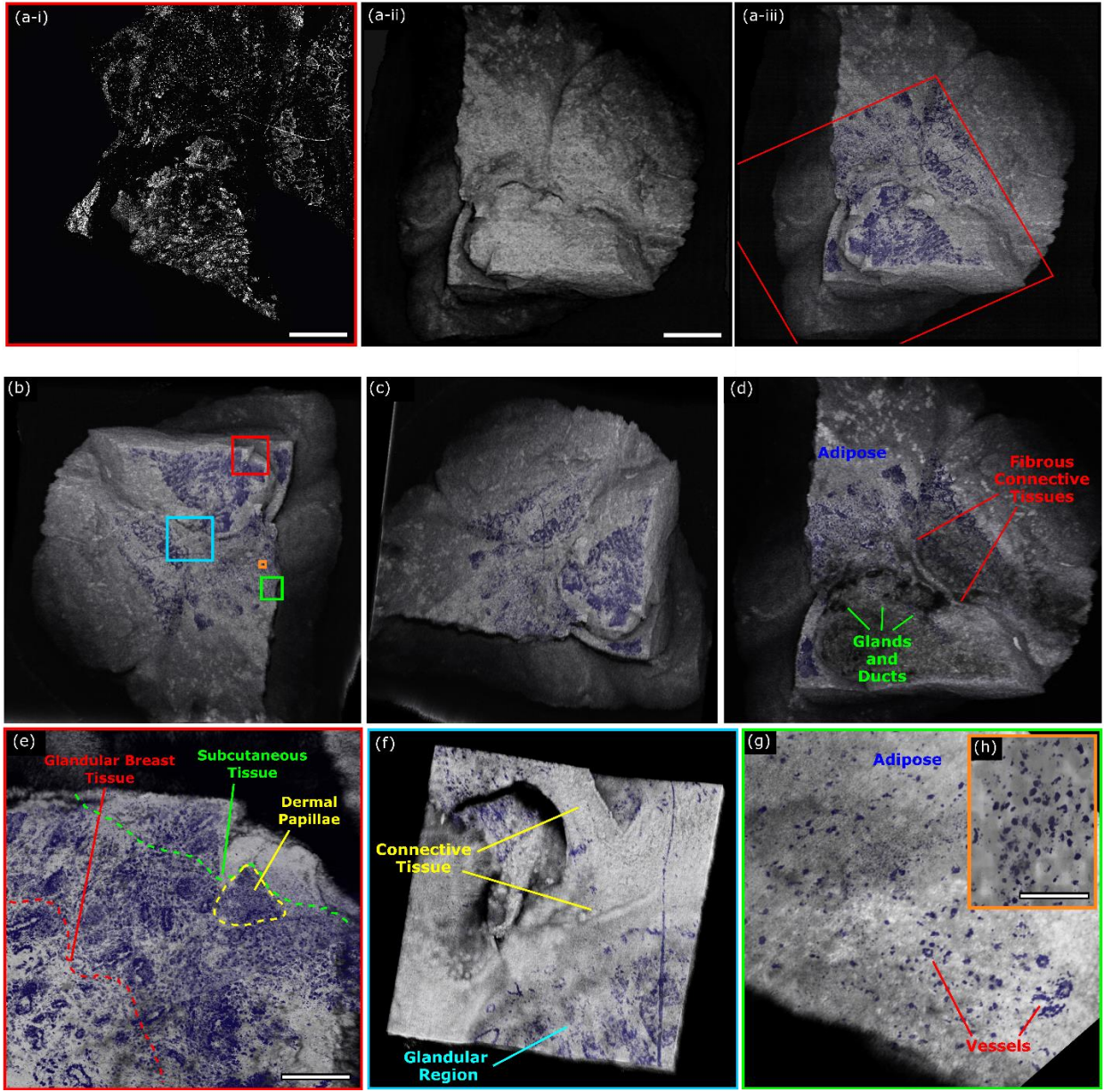

**Figure 5:** Combined PARS and OCT imaging of unprocessed excised Rattus mammary tumor tissue. (a) (i) PARS image of resected Rattus mammary tumor. Scale Bar: 1 mm. (ii) OCT (0.14 NA) image of entire section of resected Rattus mammary tumor tissues. Scale Bar: 1.25 mm. (iii) Merged PARS and OCT image of resected Rattus mammary tumor. Red box outlines limits of PARS image. (b, c) Different angles of 3-dimensional representation of merged PARS and OCT images of excised Rattus mammary tumor tissue. (d) Vertical subsection of PARS and OCT image showing subsurface tumor tissue morphology and nuclear visualizations. (e-h) Merged PARS and OCT (0.4 NA) image of resected Rattus mammary tumor. (e) Scale Bar: 250 μm. (h). Scale Bar: 100 μm.

Moving forwards, several technical challenges must be overcome to further improve the clinical functionality of the proposed system. One of the main limitations of the current PARS architecture is the 50 kHz excitation. With the existing UV source, we are limited to an imaging speed of ~ 90 seconds per mm$^2$. To reduce imaging times, a new 2.7 MHz excitation source is being implemented. The new source will reduce imaging times to a few seconds per square millimeter. This will enable real-time PARS histological imaging, and near real-time volumetric imaging. Another challenge which arises with the current implementation is the specificity of the combined contrast. While the PARS and OCT system recover nuclear morphology and tissue structures, the scattering contrast of OCT does not provide tissue specificity. Therefore, it may be difficult to differentiate extranuclear biomolecules such as collagen or lipids.

To resolve this issue, it may be advantageous to add further PARS excitation wavelengths to the proposed system. More excitation wavelengths would expand the chromophore specific absorption contrast. This may provide visualizations of biomolecules such as lipids, collagen, heme proteins and more [15,19,28,39].

Several refinements and adaptations could also be made to the OCT subsystem. The swept source OCT only provides ~7.3 µm axial and ~3.9 µm lateral resolution. This is substantially less than the ~1.4 µm axial and ~300 nm lateral resolution of the PARS system. As a result, the OCT subsystem cannot resolve structures on the same scale as the PARS microscope. This challenge is compounded when considering the PARS system may only provide volumetric imaging to ~100 µm deep[14,17]. Moving forwards, it may be advantageous to develop a higher axial resolution OCT system. If the PARS and OCT axial resolution were equal, the combined three-dimensional representations could capture co-registered depth resolved images of subcellular structures including single nuclei. Another challenge to consider is the PARS and OCT subsystems use independent objective lenses. As a result, the co-alignment between PARS and OCT images is not perfect as tissues can potentially shift when switching between objectives. In the ideal case, rather than imaging with PARS and OCT independently, both images would be captured simultaneously through the same lens. This could significantly reduce imaging time. Moreover, if the PARS and OCT images were collected together, the OCT A-lines could be used to determine the height of the tissue surface at each location. An adaptive optics, or mechanical scanning system could then adjust the PARS system focus based on the OCT A-line to guide imaging of rough tissue surfaces.

**Conclusions**

This work represents the first report of: (1) label-free histological imaging in human tissue specimens with a non-contact photoacoustic system, (2) 3-dimensional imaging of subsurface nuclear morphology in resected tissue specimen with a non-contact photoacoustic system, and (3) a combined photoacoustic and OCT microscope for label-free visualizations of nuclear and extranuclear morphology in resected tissue specimens. As presented, the proposed multimodal microscope provides co-registered sub-micron depth-resolved nuclear imaging, and micron scale depth resolved tissue morphology in an all-optical reflection mode format. Moreover, these visualizations may be recovered label-free directly from thick unprocessed tissues. The clinically relevant diagnostic features captured by the proposed system are not accessible through any other means. The unparalleled clinical diagnostic potential of this system could revolutionize the workflow for diagnosis, treatment, and resection of cancer and other neoplasms. Applied in a clinical setting, the proposed system would allow pathologists to rapidly assess subsurface nuclear structures without extensive sample processing and reduce the potential for sampling error in larger pathology specimens. Bulk tissues could undergo histological imaging immediately after excision. This would improve the diagnostic yield of histological preparations, while reducing histological imaging timelines from weeks to minutes. Moreover, this would preserve whole tissues for further histopathological processing and immunochemistry. Moving forwards, this technology will soon be applied to imaging a wider variety of resected human tissues.

**Acknowledgements**

The authors, thank Jean Flanagan at the University of Waterloo animal facility for her work in procuring Rattus tumor tissue specimens.

The authors thank the following sources for funding used during this project. Natural Sciences and Engineering Research Council of Canada (DGECR-2019-00143, RGPIN2019-06134); Canada Foundation for Innovation (JELF #38000); Mitacs Accelerate (IT13594); University of Waterloo Startup funds; Centre for Bioengineering and Biotechnology (CBB Seed fund); illumiSonics Inc (SRA #083181); New frontiers in research fund – exploration (NFRFE-2019-01012).

**Additional Information**

*Conflict of interest:*

Authors B.R. Ecclestone, K. Bell, D. Dinakaran, J.R. Mackey and P.H. Reza have financial interests in illumiSonics Inc. IllumiSonics partially supported this work. Authors Z. Hosseinaee and N. Abbasi declare no conflicts of interest

*Data Availability:*

All data generated or analyzed during this study are included in this published article.

**Author Contributions:**

B.R.E., Z.H., N.A., and K.B constructed the PARS and OCT system, and executed imaging experiments. B.R.E. wrote the main manuscript text and prepared the figures. D.D., and J.R.M., selected and processed clinical samples, and provided consultation in writing the manuscript. P.H.R., conceived the project and acted as the primary investigator.